\documentclass[11pt]{article}
\usepackage{amsbsy}
\usepackage{amssymb}
\newcommand{\beq}{\begin{equation}}
\newcommand{\eeq}{\end{equation}}
\newcommand{\bdm}{\begin{displaymath}}
\newcommand{\edm}{\end{displaymath}}
\newcommand{\beqr}{\begin{eqnarray}}
\newcommand{\eeqr}{\end{eqnarray}}
\newcommand{\beqrn}{\begin{eqnarray*}}
\newcommand{\eeqrn}{\end{eqnarray*}}
\def\l{\lambda}
\def\R{{\Bbb R}}
\def\bchi{\boldsymbol{\chi}}
\def\ba{\boldsymbol{a}}

\def\ve{\varepsilon}
\def\a{\alpha}
\def\k{\kappa}
\def\l{\lambda}
\def\nn{\nonumber}

\begin{document}

\title{Some new solutions to the Schr\"odinger equation for the trigonometric $E_8$ Calogero-Sutherland problem}

\author{J. Fern\'andez N\'u\~{n}ez$^*$, W. Garc\'{\i}a
Fuertes$^*$,
A.M. Perelomov$^\dag$}

\date{}

\maketitle
\begin{center}
$^*${{\it Departamento de F\'{\i}sica, Facultad de
Ciencias,  Universidad de Oviedo,\\
E-33007 Oviedo, Spain.}} 

$^\dag${\it  Institute for Theoretical and Experimental Physics, 117259, Moscow, Russia.}
\vskip1cm
\end{center}
\vskip1cm
\begin{abstract}
\noindent
We provide a list of explicit eigenfunctions of the trigonometric Calogero-Sutherland Hamiltonian associated to the root system of the exceptional Lie algebra $E_8$. The quantum numbers of these solutions correspond to the first and second order weights of the Lie algebra.
\end{abstract}
\vfill\eject
\section{Introduction}
Since their discovery by Calogero \cite{ca71} and Sutherland \cite{su72}, the integrable models bearing their names have been thoroughly investigated by many researchers. In particular, in the paper \cite{op76} such models were extended to the case of arbitrary root systems of semi-simple Lie algebras. Despite the many beautiful mathematical developments and interesting physical applications which arose from all this work \cite{dv00,poly}, it is rare to find in the literature explicit solutions to the Schr\"odinger equation for this class of systems. In fact, some very sound mathematical approaches have been proposed to address this problem but, to our knowledge, they have been applied in explicit form to obtain only a handful of the eigenfunctions for the most simple cases. We feel, however, that in this business it would be desirable to go one step further from the development of abstract and general mathematical methods and to present instead, once and for all, some concrete results on the eigenfunctions. Even if the formulas involved are rather unwiedly, it seems to us that they can be very useful. It is nowadays possible to handle these formulas by means of a variety of powerful symbolic computer programs, and the availability of collections of concrete eigenfunctions for the different versions of the Calogero-Sutherland problem can be a considerable help for researchers if they need to put to the test some conjectures, to check the reliability and accuracy of new computational schemes, or for several other matters in this domain. Note also that, as for particular values of the coupling constant the eigenfunctions represent remarkable mathematical objects, such a characters of Lie algebras or zonal spherical functions on symmetric spaces, their computation and knowledge is interesting also from the point of view of pure mathematics.

In \cite{pe98a}, one of us put forward an approach to the diagonalization of the trigonometric Calogero-Sutherland Hamiltonian with $SU(3)$ root system which exploits its Weyl invariance by using as dynamical variables the characters of the fundamental irreducible representations of the $SU(3)$ algebra. This procedure yields a second order differential operator in these variables whose coefficients are polynomials over the integers. Being polynomial, the differential equation can be solved by an iterative method, and in this way it is possible to find not only the wave functions in explict form, but also some other noteworthy results such as recurrence relations among the eigenfunctions, generating functions for particular sets of solutions of the Schr\"{o}dinger equation, lowering and raising operators connecting the energy levels, etc. The approach has been subsequently developed in a series of papers devoted to the Calogero-Sutherland models associated to the root systems of $SU(n)$ \cite{sun}, $SO(8)$ \cite{so8}, and the exceptional $E_6$, $E_7$ and $E_8$ series \cite{e678, hame8}, and has given many results of the aforementioned type for all these cases. The last paper of the series appeared quite recently and was devoted to obtain the Hamiltonian in Weyl-invariant variables for the case of the Lie algebra $E_8$. Nevertheless, due to their length, the paper was not a good place to present examples of eigenfunctions, except for a few of the simplest ones. Thus, we publish now this preprint to make available to the researches interested in the field a more complete list of solutions of the Schr\"{o}dinger equation for the $E_8$ case.
\section{Review of the theory}
The  trigonometric Calogero-Sutherland model related to the root system
$\cal R$ associated to a simple Lie algebra of of rank $r$ is the quantum system in the Euclidean space $\R^r$ describing a set of parcicles moving in a circle,  
defined by the standard Hamiltonian operator
\beq
\label{ham}
H=\frac{1}{2}\sum_{j=1}^rp_j ^2+\sum_{\a\in{\cal
R}^+}\kappa_\a(\kappa_\a-1)\sin^{-2}(\a,q),
\eeq
where $q=({q_j})$ is  the Cartesian coordinate system provided by the canonical basis  of $\R^r$ and $p_j=-{\rm
i}\,\partial_{q_j}$, and $(\cdot,\cdot)$ is the standard scalar product in $\R^r$; ${\cal R}^+$  is the set of the positive roots of $L$,
and the coupling constants $\k_\a$ are such that $\kappa_\a=\kappa_\beta$ if
$|\a|=|\beta|$. We are interested only in the case of simply-laced root
systems (as the $E$-series is),  for which the Calogero-Sutherland model
depends only on one coupling constant $\k$.

To find the stationary states, it is necessary to solve the
Schr\"odinger eigenvalue problem $H\Psi=E\Psi$. The following important
facts about this family of  quantum mechanical systems were   established
in \cite{op83}.

(a)The ground state energy and (non-normalized) wave function of these integrable systems are
\begin{eqnarray}
E_0(\k)&=&2 \rho^2\k^2\nn\\
\Psi_0^\k(q)&=&{\prod_{\a\in {\cal R}^+}\sin^\k(\a, q)},
\end{eqnarray}
with $\rho$ being the Weyl vectorç
\beq
\rho=\frac{1}{2}\sum_{\a\in\cal R^+}\a 
\eeq
of the algebra, while the excited states are indexed by the highest weights $\mu=\sum
m_i\l_i\in P^+$ (where $P^+$ is the cone of dominant weights) of the irreducible
representations of $L$, that is, by a $r$-tuple of non-negative integers
${\bf m}=(m_1,\dots,m_r)$ (the quantum numbers). The wave functions $\Psi_{\bf m}^\k$ and the energy levels $E_{\bf m}(\k)$ satisfy
\begin{eqnarray}
H\Psi^\k_{\bf m}&=&E_{\bf m}(\k)\Psi_{\bf m}^\k\nn\\
E_{\bf m}(\k)&=&2 (\mu+\k\rho,\mu+\k\rho).\label{105}
\end{eqnarray}

(b)It is natural to look for the solutions $\Psi_{\bf m}^\k$ in the form
\beq
\Psi_{\bf m}^\k(q)=\Psi_0^\k(q)\Phi_{\bf m}^\k(q),
\eeq
and consequently we are led to the eigenvalue problem
\beq
\Delta^\k\Phi_{\bf m}^\k=\ve_{\bf m}(\k)\Phi_{\bf m}^\k\,,
\label{sch}
\eeq
where $\Delta ^\k$ is the linear differential operator
\beq
\Delta^\k=-\frac{1}{2}\sum_{j=1}^r\partial_{q_j}^{\,2}-\k\sum_{\a\in {\cal
R}^+}  {\cot}(\a, q)(\a,\partial_q)
\label{dk},
\eeq
and the eigenvalues $\ve_{\bf m}(\k)$ are the energies over the ground
level, i.e.,
\beq
\label{energ}
\ve_{\bf m}(\k)=E_{\bf m}(\k)-E_0(\k)= 2(\mu, \mu+2\k\rho).
\eeq
(c)In the case $\k=0$ the wave functions (\ref{sch}) are (proportional to)
the monomial symmetric functions
\beq
M_\l(q)=\sum_{w\in W}e^{2i(w\cdot \l,q)},\ \l\in P^+\,,
\eeq
$W$ being the Weyl group of $L$. And the wave functions in the case $\k=1$
are (proportional to) the characters of the irreducible representations
\beq
\label{bchi}
\bchi_\l(q)=\frac{\sum_{w\in W}(\det w)e^{2i(w\cdot(\l+\rho),q)}}{\sum_{w\in
W}(\det w)e^{2i(w\cdot \rho,q)}},\ \l\in P^+\,.
\eeq
Both $M_\l$ and $\bchi_\l$ are sums over the orbit $\{w\cdot\l\}$ of $\l$ under $W$, and
consequently, $W$-invariant; as wave functions, they represent
superpositions of plane waves. 

\medskip
Due to the Weyl symmetry of the Hamiltonian, the wave functions
$\Phi_{\bf m}^\k(q)$ are   $W$-invariant, and the best way to solve the
eigenvalue problem (\ref{sch}) is to use the set of independent
$W$-invariant variables $z_k=\bchi_{\l_k}(q)$, in terms of which the wave
functions $\Phi_{\bf m}^\k$ are polynomials.
Although the expression of these characters $z_k$ in terms of the
$q$-coordinates is very complicated, it is possible to perform the change of variables by an indirect route, as explained in \cite{hame8}. The $\Delta^\k$ operator takes then the form
\beq
\Delta^\kappa=\sum_{j\leq k} \ba_{jk}(z)\partial_{z_j}\partial_{z_k}+\sum_{j}
\left[b_j^1(z)+(\kappa-1)b_j(z)\right]\partial_{z_j},
\label{deltaz}
\eeq
and the coefficients were found in \cite{hame8}. Once the operator $\Delta^\k$ is known, the eigenfunctions can be calculated by an iterative procedure, see the for instace \cite{so8}.
\section{First and second order polynomials}
In this paper, we provide a list of the polynomials $\Phi_{\bf m}^\k(z)$ such that  ${\displaystyle \sum_{k=1}^8 m_k=1\;{\rm or}\;2}$, i.e. whose highest weight term is linear or quadratic in the $z$-variables. Nevertheless, because these polynomials are quite long, it does not seem very useful to publish them as a part of the preprint itself, but instead as separate text files which can easily be copied to be used in symbolic programs such as Mathematica or Maple. Thus, along with this main body of the preprint, we are submitting to the arXives three independent files under the names hamiltonian.txt, orderone.txt and ordertwo.txt. The first of these files contains the coefficients of the Hamiltonian such as they are described and listed in \cite{hame8}. Notice that the coefficients $\ba_{jk}(z)$, $b_j^1(z)$ and $b_j(z)$ are written in the text file as a[j,k], bj1 and bj, respectively. The polynomials which solve the Schr\"{o}dinger equation are placed into the other two files. In them, the character ``x" denotes the coupling constant $\kappa$ of the Calogero-Sutherland potential, and the notation P[1,0,0,0,0,0,0,0] is for the polynomial $\Phi_{\bf m}^\k(z)$ for ${\bf m}=(1,0,0,0,0,0,0,0)$. The procedure to get the text files is to download the preprint in source format and decompress it with the appropriate sofware (see the instructions in the arXives webpage).

\section*{Acknowledgements} This paper was completed during the visit of one
of the authors
(AMP) to the Max-Planck-Institut f\"ur Gravitationsphysik,
and he thanks the institute staff for the hospitality. This work was partially supported by  Spanish Government under grants MTM2006-10532 (JFN) and FIS2006-09417 (WGF and AMP).

\end{document}